# Quantum Logic Operations Using Single Photons and the Zeno Effect

J.D. Franson, B.C. Jacobs, and T.B. Pittman

Johns Hopkins University

Applied Physics Laboratory

Laurel, MD 20723

*Abstract:*

We show that the quantum Zeno effect can be used to implement several quantum logic gates for photonic qubits, including a gate that is similar to the square-root of SWAP operation. The operation of these devices depends on the fact that photons can behave as if they were non-interacting fermions instead of bosons in the presence of a strong Zeno effect. These results are discussed within the context of several no-go theorems for non-interacting fermions or bosons.

In the quantum Zeno effect [1, 2], a randomly-occurring event can be suppressed by frequent measurements to determine whether or not the event has occurred. Here we show that the Zeno effect can be used to implement quantum logic operations that require only linear optical elements, aside from the Zeno effect itself. The operation of these devices depends on the fact that photons can behave in some respects as if they were non-interacting fermions instead of bosons in the presence of a strong Zeno effect. No actual observations are required and the same results can be obtained, at least in principle, using two-photon absorption to record the presence of more than one photon in an optical fiber.

The implementation of Zeno gates of this kind would have a number of potential advantages for an optical approach to quantum computing [3-7], but the feasibility of that approach is beyond the intended scope of this paper. Here we describe some of the interesting theoretical aspects of photons subjected to a strong Zeno effect, including the implications of fermionic behavior for various no-go theorems [8-12] in quantum computing and the reversal of the well-known Hong-Ou-Mandel interference pattern [13]. A discussion of the practical issues involved in the development of Zeno gates will be submitted for publication elsewhere.

The most general quantum computation can be performed using single-qubit logic gates, which can be easily implemented in an optical approach, combined with a suitable two-qubit gate, such as the controlled-NOT [14]. The most useful two-qubit gate for our purposes is the $\sqrt{SWAP}$ gate [15, 16] illustrated in Figure 1a. The SWAP operation itself interchanges the logical values of two input qubits, and it could easily be implemented in a number of ways. The $\sqrt{SWAP}$ gate is defined as giving the SWAP operation when applied twice (or squared), as illustrated in the figure. This is a purely quantum-mechanical operation and its implementation is nontrivial. The $\sqrt{SWAP}$ allows any quantum computation to be performed when combined with single-qubit operations, and it can be used to implement the more familiar controlled-NOT gate [15].

A potential implementation of a $\sqrt{SWAP}$ gate for photons is illustrated in Figure 1b. Here the cores of two optical fibers are brought into close proximity over a distance $L$ as further illustrated in Figure 2a. Photons in each fiber are assumed to occupy only the fundamental spatial mode with a fixed linear polarization. A photon in one fiber core will be weakly coupled into the other by their evanescent fields, as illustrated by the dotted arrows in Figure 1. If $L$ is chosen properly, a photon incident in one of the fiber cores will be completely transferred to the other fiber core. This process will implement the SWAP operation if we take the presence of a photon to represent the logical value 1 and the absence of a photon to represent the logical value 0. The basic idea is to implement the $\sqrt{SWAP}$ by shortening the length of the interaction region to a length $L_{1/2} = L/2$, as illustrated by the dashed line in the middle of Figure 1b. This will produce an operation that, when applied twice, might be expected to give the SWAP operation as desired. Commercially-available devices with similar structures are used as 50/50 couplers and are equivalent to free-space beam splitters.

It can be shown that the coupled-fiber device of Figure 1b does implement a $\sqrt{SWAP}$ operation correctly if a total of 0 or 1 photons are input to the device. Incorrect results are obtained, however, if a photon is present in both of the input modes, which corresponds to a logical input of 1 for both qubits. In that case, there is some probability that both photons will exit the device in the same fiber core, which corresponds to an error state since only 0 or 1 photons represents a valid logical output.

In fact, quantum interference effects ensure that both photons will always exit such a device in the same fiber core, which is equivalent to the well-known Hong-Ou-Mandel dip [13] in coincidence measurements using free-space beam splitters.

In order to investigate the ability of the quantum Zeno effect to inhibit the production of these kinds of errors, we assumed that $N$ equally-spaced measurements to determine whether or not two photons are present in the same fiber core are performed during the time interval $\Delta t = L_{1/2}/c$ that the photons interact, where $c$ is the speed of light. Between measurements, the system was assumed to undergo unitary evolution governed by the Hamiltonian $\hat{H}$ given by

$$\hat{H} = \hbar\omega(\hat{a}_1^\dagger \hat{a}_1 + \hat{a}_2^\dagger \hat{a}_2) + \varepsilon(\hat{a}_1^\dagger \hat{a}_2 + \hat{a}_2^\dagger \hat{a}_1) \qquad (1)$$

Here $\hbar$ is Planck's constant divided by $2\pi$, $\omega$ is the angular frequency of the photons, the operators $\hat{a}_1^\dagger$ and $\hat{a}_2^\dagger$ create photons in each of the fibers, and the parameter $\varepsilon$ determines the strength of the coupling between the two fibers. The device was assumed to fail if a measurement determined the presence of two photons, whereas the absence of two photons projected the state vector into the orthogonal subspace. It should be noted that here we are observing the presence of the system in a particular final state, whereas the original Zeno effect [1] involved an observation of the system in its initial state. A more detailed description of this calculation and those that follow will be submitted for publication elsewhere [17].

The effect of such a sequence of measurements is illustrated by the dots in Figure 3, which shows the probability $P_E$ that an incorrect result will be produced by the device. It can be seen that $P_E = 1$ if a single measurement is made at the end of the process, which is consistent with the Hong-Ou-Mandel dip mentioned above. But $P_E$ is inversely proportional to $N$ in the limit of large $N$, as would be expected from the quantum Zeno effect, and the output of the device can be shown to be that of a $\sqrt{SWAP}$ gate in that limit aside from a phase factor described below. As a practical matter, failure events in which two photons are detected in the same fiber can be corrected [17] using the two-qubit encoding introduced by Knill, Laflamme, and Milburn [3].

To be more specific, such a device implements a unitary transformation given by

$$\sqrt{SWAP'} = \begin{bmatrix} 1 & 0 & 0 & 0 \\ 0 & (1+i)/2 & (1-i)/2 & 0 \\ 0 & (1-i)/2 & (1+i)/2 & 0 \\ 0 & 0 & 0 & i \end{bmatrix} \qquad (2)$$

Here a fixed phase shift of $\exp(i\pi/4)$ was included in each of the output ports in order to facilitate comparison with the usual definition of the $\sqrt{SWAP}$ operation. It can be seen that Eq. (2) differs from a conventional $\sqrt{SWAP}$ by a factor of $i$ in the lowest diagonal element. As a result, the $\sqrt{SWAP'}$ operation has the property that its square, which we will denote by $SWAP'$, has the form

$$SWAP' \equiv (\sqrt{SWAP'})^2 = \begin{bmatrix} 1 & 0 & 0 & 0 \\ 0 & 0 & 1 & 0 \\ 0 & 1 & 0 & 0 \\ 0 & 0 & 0 & -1 \end{bmatrix} \qquad (3)$$

It can be seen that $SWAP'$ differs from the conventional definition of a $SWAP$ operation by the factor of $-1$ in the lowest diagonal element, and that a controlled-Z operation (controlled Pauli z-matrix) can be implemented by applying $SWAP'$ followed by $SWAP$.

In the case of photonic qubits in optical fibers, the *SWAP* operation can be applied by simply crossing the optical fibers as illustrated in Fig. 4a, which allows a controlled-Z operation to be implemented as shown in Figure 4b. A controlled-NOT gate can then be implemented by applying a Hadamard transformation [14] (a simple beam splitter operation) to the target qubit before and after the circuit shown in Fig. 4b. Dual-rail techniques can be used to avoid the need to create or annihilate photons during the Hadamards.

The quantum interference effects responsible for the Hong-Ou-Mandel dip and the value of $P_E = 1$ in the absence of the Zeno effect (left-hand-side of Fig. 3) are due to the fact that photons are bosons [18, 19]. Electrons or other fermions would give just the opposite result, with both particles always exiting from different output ports [18, 19]. This difference in behavior can be traced to the fact that the exchange of two identical fermions multiplies the state vector by a factor of $-1$, whereas the exchange of two bosons gives a factor of $+1$, which converts an interference maximum to a minimum. It can be seen from Figure 3 that the properties of the photons, at least as far as these interference effects are concerned, gradually change from that of a boson to that of a fermion as the strength of the Zeno effect is increased. A strong Zeno effect clearly prevents two photons from occupying the same mode, which is a fundamental property of fermions (the Pauli exclusion principle). Whether or not this behavior continues to hold true for other properties of the photons, such as their thermodynamic behavior, remains to be investigated.

The fermionic behavior of the photons can be further understood by comparing the results of the above calculation with the corresponding calculation for non-interacting electrons (with no measurements or Zeno effect). Here the relevant

Hamiltonian is still given [8] by Eq. (1) while its matrix elements between the various basis states can now be calculated using the anti-commutation relations for fermions, such as $\{\hat{a}_i^\dagger, \hat{a}_j^\dagger\} = 0$. For a single incident particle, there is no difference between using commutation or anti-commutation relations and the matrix elements are the same as for an incident boson. The case of one incident fermion in each path is also straightforward because there is no coupling to any other state, as is also the case for two incident photons in the presence of a strong Zeno effect. As a result, the device of Fig. 1b produces exactly the same operation for non-interacting fermions as it does for photons in the presence of the Zeno effect, namely the $\sqrt{SWAP'}$ of Eq. (2). This shows that the behavior of photons in the presence of a strong Zeno effect at a beam splitter is the same as that of non-interacting fermions, including all phase shifts and sign reversals. (This result is obviously restricted to the case in which at most one photon is incident in each mode, which is enforced by the Zeno effect itself.)

This raises the interesting question as to whether or not the circuit shown in Fig. 4b could be used to implement a controlled-Z and thus a controlled-NOT gate for non-interacting fermions, such as electrons. Due to the anti-commutation relations, interchanging the paths of two electrons as in Fig. 4a will implement the $SWAP'$ operation instead of $SWAP$, and the circuit of Fig. 4b would only produce the identity operation for the case of fermions. This is consistent with various "no-go" theorems [8, 9] that show that universal quantum computation cannot be implemented using non-interacting fermions. The ability to perform quantum computation using single photons and the Zeno effect is due to the fact that the photons can behave as non-interacting fermions in one part of a circuit and as non-interacting bosons in other parts of the circuit, which avoids the no-go theorems for either non-interacting fermions [8, 9] or bosons [11, 12]. These no-go theorems can also be avoided in probabilistic logic operations by using ancilla and feed-forward control based on the results of measurements [3-7, 10].

We would now like to return to the question of how Zeno gates of this kind could be implemented, at least in principle. As is usually the case in the quantum Zeno effect, it is not necessary to make any actual observations or measurements as described above. Instead, it suffices to introduce an interaction between the system of interest and some other system whose properties could, in principle, be measured at some later time in order to obtain the same information. In the situation of interest here, that can be accomplished by introducing one or more atoms into the fiber cores in such a way that the atoms can absorb two photons but not just one [20]. In order to investigate this possibility, we assumed that the quantum states corresponding to two photons in the same fiber core would undergo two-photon absorption at a rate of $1/\tau_D$, where $\tau_D$ is the corresponding decay time. The solid line in Fig. 3 illustrates the results of a density-matrix calculation of this kind. In order to allow both sets of calculations to be plotted on the same scale, the parameter $N$ was defined in this case as $N = \Delta t / 4\tau_D$. It can be seen that strong two-photon absorption can also suppress the emission of two photons into the same mode, thereby preventing errors of that kind. The Zeno effect has previously been considered for use in quantum error correction [21-23] and in a coherent-state approach [24].

Nonlinear effects such as two-photon absorption are commonly assumed to be small at single-photon intensities. A simple argument illustrated in Fig. 2b, however, suggests that two-photon absorption can be substantial if the photons are confined inside a hollow optical fiber containing a suitable atomic vapor, for example. The absorption of a resonant photon by an atom can be described by a cross-sectional area that is comparable to the area of an optical fiber core [25] in the absence of collisions. A single photon will then be absorbed by a single atom in the fiber core with a probability that is on the order of unity. Once one photon has been absorbed, the atom will be left in an excited state with a similar cross-section, so that a second photon can then be absorbed with a probability that is also on the order of unity. The absorption of a single photon can be avoided by detuning from the resonant frequency of the atomic transition. Although the effects of collisions, Doppler shift, and detuning must also be included in any analysis, this argument does suggest that two-photon absorption can occur at a relatively high rate in optical fibers under the right conditions. It may also be possible to obtain similar effects using solid-core fibers doped with suitable atoms. An analysis of practical considerations of this kind will be discussed elsewhere [17].

In summary, we have shown that the quantum Zeno effect can be used to implement several types of quantum logic gates for single photons. No actual measurements are required and two-photon absorption in dual-core optical fibers would suffice for the construction of a $\sqrt{SWAP'}$ gate. The operation of these devices depends on the fact that photons can behave as non-interacting fermions in the presence of a strong Zeno effect in one part of a circuit and as non-interacting bosons in other parts of the circuit, which avoids various no-go theorems for non-interacting particles [8-12].

We would like to acknowledge discussions with C.W.J. Beenakker, Mark Heiligman, Tim Ralph, and Colin Williams. This work was supported in part by ARO, ARDA, NSA, ONR, and IR&D funds.

**References:**

1. B. Misra and E.C.G. Sudarshan, E.C.G, J. Math. Phys. **18**, 756 (1977).

2. A.G. Kofman and G. Kurizki, Phys. Rev. Lett. **87**, 270405 (2001).

3. E. Knill, R. Laflamme, and G.J. Milburn, Nature (London) **409,** 46 (2001).

4. T.B. Pittman, B.C. Jacobs, and J.D. Franson, Phys. Rev. A **64**, 062311 (2001).

5. T.B. Pittman, B.C. Jacobs, and J.D. Franson, Phys. Rev. Lett. **88**, 257902 (2002).

6. J.D. Franson, M.M. Donegan, M.J. Fitch, B.C. Jacobs, and T.B. Pittman, Phys. Rev. Lett. **89**, 137901 (2002).

7. T.B. Pittman, M.J. Fitch, B.C. Jacobs, and J.D. Franson, Phys. Rev. A **68**, 032316 (2003).

8. B.M. Terhal and D.P. DiVincenzo, Phys. Rev. A **65**, 032325 (2002).

9. E. Knill, quant-ph/0108033.

10. C.W.J. Beenakker, D.P. DiVincenzo, C. Emary, and M. Kindermann, quant-ph/0401066.

11. L. Vaidman and N. Yoran, Phys. Rev. A **59**, 116 (1999).

12. N. Lutkenhaus, J. Calsamiglia, and K.-A. Suominen, Phys. Rev. A **59**, 3295 (1999).

13. C.K. Hong, Z.Y. Ou, and L. Mandel, Phys. Rev. Lett. **59**, 2044 (1987).

14. M.A. Nielsen and I.L. Chuang, *Quantum Computation and Quantum Information* (Cambridge U. Press, Cambridge, 2000).

15. D. Loss and D.P. DiVincenzo, Phys. Rev. A **57**, 120 (1998).

16. K. Eckert, J. Mompart, X.X. Yi, J. Schliemann, D. Bruss, G. Birkl, and M. Lewenstein, Phys. Rev. A **66**, 042317 (2002).

17. J.D. Franson, B.C. Jacobs, and T.B. Pittman, to be submitted to Phys. Rev. A.

18. R. Loudon, Phys. Rev. A **58**, 4904 (1998); H. Fearn and R. Loudon, J. Opt. Soc Am. B **6**, 917 (1999).

19. Although photon "anti-bunching" can also be observed using entangled polarization states, that does not imply fermionic statistics. Here we assume that all the photons are in the same spatial and polarization mode.

20. R.W. Boyd, *Nonlinear Optics*, *2$^{nd}$ edition* (Academic Press, New York, 2003).

21. W.H. Zurek, Phys. Rev. Lett. **53**, 391 (1984).

22. A. Barenco, A. Berthiaume, D. Deutsch, A. Eckert, R. Josza, and C. Macchiavello, S.I.A.M. Journal on Computing **26**, 1541 (1997).

23. L. Vaidman, L. Goldenberg, and S. Wiesner, Phys. Rev. A **54**, R1745 (1996).

24. T.C. Ralph, A. Gilchrist, G.J. Milburn, W.J. Munro, and S. Glancy, Phys. Rev. A **68**, 042319 (2003).

25. L. Mandel and E. Wolf, *Optical Coherence and Quantum Optics* (Cambridge U. Press, Cambridge, 1995).

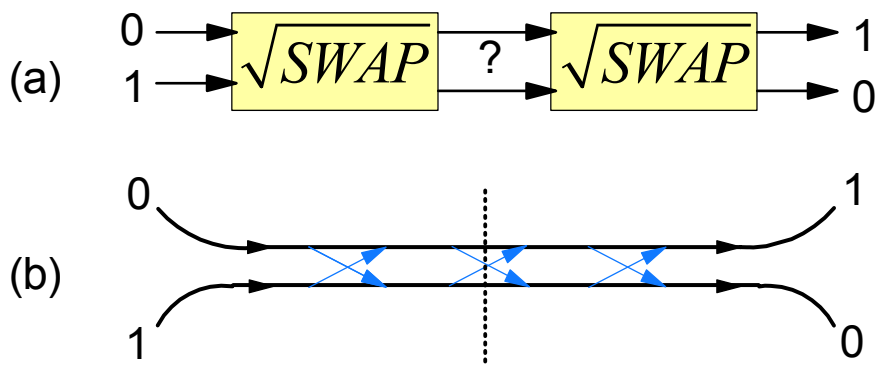

Figure 1. (a) Operation of the $\sqrt{SWAP}$ gate, which is defined in such a way that it swaps the values of any two logical inputs when it is applied twice (squared). If the operation is only applied once, however, a quantum superposition of states is created and neither qubit has a well-defined value, as illustrated by the question mark. (b) Potential implementation of a $\sqrt{SWAP}$ gate by bringing the cores of two optical fibers in close proximity, which allows a photon in one core to be coupled into the other core. If length L produces a SWAP operation, then half that length (dashed line) will produce the $\sqrt{SWAP}$ operation, aside from errors that can be suppressed using the Zeno effect and a phase factor discussed in the text.

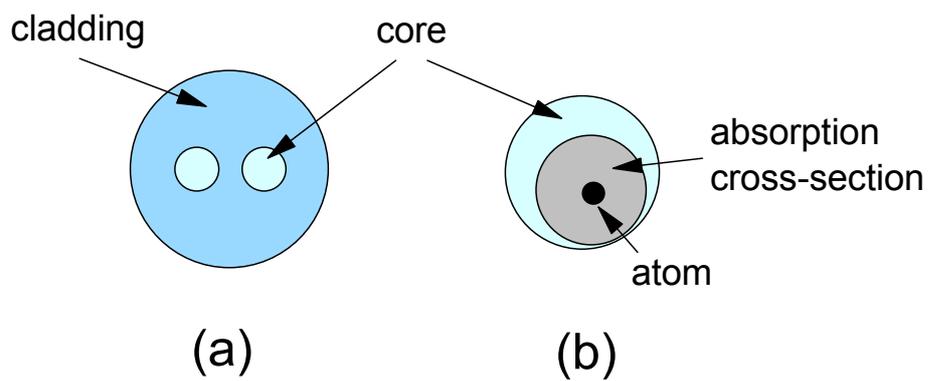

Figure 2. (a) End view of a double-core optical fiber used to implement a $\sqrt{SWAP'}$ operation as in Figure 1b. (b) Expanded view of one of the optical fiber cores containing a single atom. The cross-section for absorption of a resonant photon can be comparable to the area of the core itself, in which case there can be a large interaction between two photons and a single atom.

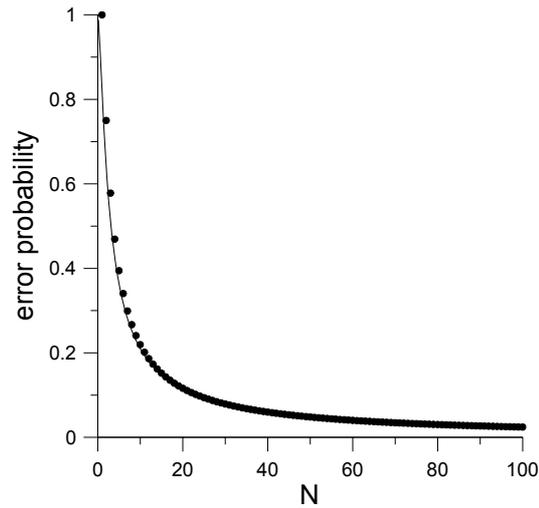

Figure 3. Probability $P_E$ of an error in the output of the $\sqrt{SWAP'}$ gate when a photon is incident in both input modes. The dots correspond to the value of $P_E$ as a function of the number $N$ of equally-spaced measurements made to determine the presence of two photons in the same optical fiber core. The solid line corresponds to similar results obtained in the presence of two-photon absorption, where the parameter $N$ is then defined as $N = \Delta t / 4\tau_D$, with $\Delta t$ the interaction time and $\tau_D$ the two-photon decay time.

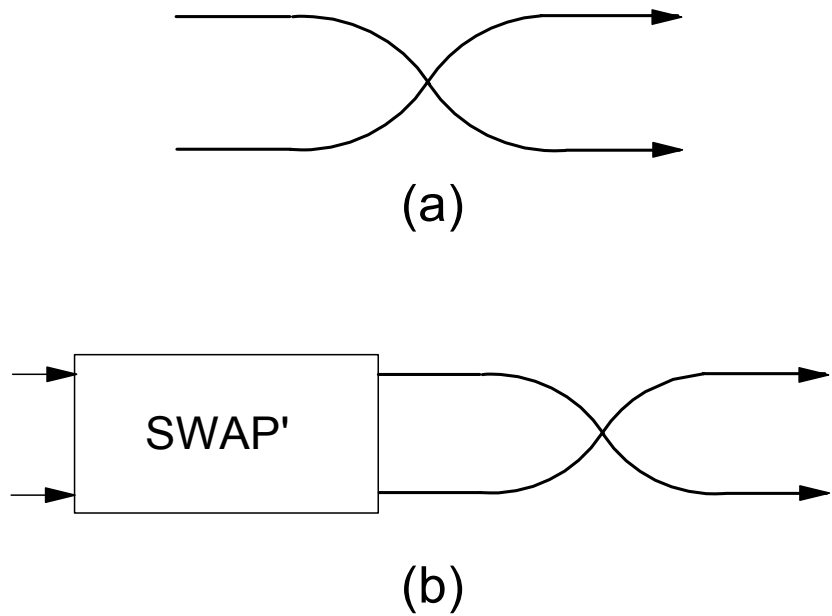

Fig. 4. (a) Implementation of a *SWAP* operation for photons by simply crossing two optical fibers. (b) A controlled-Z gate for photons constructed by applying the *SWAP'* operation of Eq. (3) followed by the *SWAP* operation illustrated above. This circuit produces only the identity operator for non-interacting fermions.